# A Perspective Study on Content Management in E-Learning and M-Learning

Dr. RD.Balaji,
Assistant Professor,
College of Applied Sciences – Salalah, Oman.
balaji.sal@cas.edu.om

Fatma Ali Al-Mahri,
Assistant Lecturer,
College of Applied Sciences –Salalah, Oman.
fatmam.sal@cas.edu.om

Malathi Balaji,
Research Scholar,
Madurai Kamaraj University, India.
bmalathisai@gmail.com

*Abstract*-**This is the era of Information and Communication Technology (ICT). Nowadays, there is no limit to learn; people can learn "anywhere and anytime" with the enhancement of technology. Electronic Learning (E-learning) and Mobile Learning (M-learning) are the two vital buzz terms in modern education particularly in Education Enhanced Technology and Technologies Supported Learning. E-learning is defined as the "instructional content or learning experience delivered or enabled by electronic technologies" whereas, M-learning is defined simply as learning via mobile devices such as cell phones, smart phones, palmtops, and handheld computers. There are many similarities between the two technologies as both are modern learning tools. Moreover, the latter is an extension and a subset of the former. However, there are few limitations or differences still exist in mobile learning tools, especially in the design, development and the technology usability. In this paper we have mainly focused on how the digital content is administrated (Content Management) in these two technologies. Additionally, the content management in E and M – learning are compared and their similarities and differences are figured out.**

*Keywords: E-Learning; M-Learning; ICT; Content Management; Technologies Supported Learning*

## I. INTRODUCTION

The Traditional Learning (T-Learning) is referred as face-to-face learning, which is commonly used type of learning, where the instructors provide learning materials to the learners in the classroom or lecture room. All the learners must be present there in person to obtain the required knowledge [1][2][3]. The instructor is responsible for transferring the knowledge to the learners, providing them with learning materials such as textbooks and lecture notes and for assessing the student's understanding during and at the end of the course by some comprehensive exams. Instructors motivate the students during their study and control the learning process [4][5] [6]. There are some limitations of this type of learning such as lack of flexibility as the learning is done in a particular place and exact time. Also, the accessibility of the learning materials is limited [6]. In addition to that, the learners must be present physically during the learning process, moreover, it is not appropriate for various other types of learning and teaching techniques; some of the well-known examples of this are online exams, discussion boards, collaboration, animation, video, listening and knowledge searching [1][2][3].

To overcome the limitations of the Traditional leaning (TL), Distance Learning (DL) programs are provided to those people who are really interested to learn but are unable to study formal education. It may be difficult for them to present physically in the academic institution and meet the instructors face-to-face [7]. DL is perfect for disabled people or those who are committed to a particular job and have no opportunity to neither get a scholarship nor study- release to study a full-time or part-time course. The interest and expectations of the learners made the researchers to bring out a new type of learning that has been developed with the help of ICT and availability of electronic media and it is referred to as Electronic Learning (E-learning) [8].

Electronic Learning (E-learning) sometimes is called online learning or an online course [4]. E-learning emerged as a new version of Distance Learning [9] and it offers learning regardless of time or place once the connection to the internet is available [4]. Using E-learning, the interactivity and efficiency of learning increased as it gives the learners higher potential to communicate more with their instructors, other learners and to access more the learning materials [5][6].

E-learning originated in the late 80s and 90s [10]. Desktops and laptops are the primary devices for e-learning [11]. The limitation of this form of study is that, E-learning through desktop restricts the learner to a specific place during the learning. Recently, many vital developments have been noticed in the mobile technologies domain particularly in laptops, notebooks, mobile phone, smart phones, wireless technology, GPRS technology and Bluetooth [12] and at the same time, Mobile devices like PDAs and cell phones which are the primary devices for M-learning [11], have become reasonable in terms of cost and functionalities and available everywhere as a result of high demand requests for trading, communications and games. Consequently, a new form of learning appears which is done via mobile devices, so the learning can be done anywhere at any time once the network is available and it is called Mobile learning [10]. With the progress and development of Information and Communication Technology (ICT), Mobile Learning (M-learning) becomes a very common type of learning [9]. M-learning is defined as gaining the knowledge or skills via mobile device technology instantly and at any place [13]. Larger number of learners uses mobile devices to write notes, read books, chat with their classmate while traveling via bus, train or even cars and even while sitting in the resultants, café [14] or health centre.

The percentage of mobile devices nowadays is 28% of the populations worldwide [15]. In Japan, in December 2004, the total number of mobile phone contracts reached approximately two third of their population [16]. A lot of changes have been noticed to the development of mobile devices over the last 10 years; incredible swift from analogy and simple cell phone to digital 3G smart phone [17]. Current mobile devices are used to transfer various types of files including text, audio and video and these smart devices work as telephone, camera, mini-computer, and PDA [17]. M-learning is considered as a "characterized technology" with its own terms unlike E-learning which has similar terminology to Traditional Learning (TL) [9].

*A. Advantages of E-learning*

Introducing e-learning into the academic institutions such as schools and colleges provides many benefits and advantages [18][19][20]:

- Flexible learning: Considering the time and location of the learning. Each learner may choose the perfect time and location to study using E-learning methods. This provides more flexibility to the learners as well as the academic institutions to provide the knowledge to their students [21].

- Effective knowledge and competence: Learners are able to have access to huge amount of knowledge easily in a minimum time which enhances their knowledge and expertise.

- Eliminates communication barriers between learners themselves and with their instructors especially for those having fear of talking to other strangers [22].

- Cost-effective: learners are not obligated to travel to gain the knowledge. Moreover, academic institutions can maximize the number of students enrolled without having to construct a new building.

- Customized learner's needs: each student seeks different knowledge based on their skills and background.

- Bridge the lack of academics and technicians.

- Self-pacing: learner can control his/her learning process depends on his/her own learning speed. As a result, learner will be more satisfied and less stress [18][20] [22] [23][ [24].

*B. Disadvantages of E-learning*

Despite E-learning advantages, there are some disadvantages found in the many studies [9][25][18][19] [21] [26][27][28][29]

- Remoteness and lack of interaction: the learner needs higher level of contemplation and time-management skills to in order to eliminate such effects.

- Low level of learning efficiency: interpretation and clarification are more effective and easier in face-to-face learning.

- Low level of learner's communication skills: learners gain much knowledge through such learning methods (E-learning) but it might be difficult to transfer the gained knowledge to other learners due to lack of such skills.

- Untrusted result of e-learning assessment: difficult to control cheating during the assessment with potentials of a high chance of plagiarism and piracy that can be done through the system. A simple way is by doing copy and paste.

- Not suitable method to all academic fields: it is perfect to learn social sciences and humanities, but not for purely scientific fields such as medical sciences and pharmacy.

- Waste of time and money: as a result of extensive use of particular websites.

II. M-LEARNING

Mobile Learning is referred to as "M-learning" or "handheld learning", a lot of research has been done on educational technology especially from technical and pedagogical perspective [30][31]. Other research discusses the limitation of mobile phone learning from psychological and pedagogical views. M-learning considered to be an extension of e-learning [32] or a new version of E-learning promoted by the use of mobile devices [33] such as cell phone, Personal Digital Assistant (PDA), smart phone, Notebook, etc. over a wireless network[34]. M-learning is defined as "the use of mobile technologies for learning"[33]. The way of learning has been changed from classroom learning to E-learning and nowadays using the mobile-learning. "Mobile-learning can be considered as and intersection of online learning and mobile computing"[34].

M-learning allow the interaction between the learner and instructor anywhere and anytime [32], m-learning has many features over E-learning. It permits the learner movement from a context to another [32] [33]. Moreover, it simplifies accessing to the course materials and activities. Also it enhances the user with personalized learning environment [6]. On the other hand, there are some constrains facing the use of mobile technology such as low resolution, small screen size, limited battery, low memory capacity, and availability of wireless network[33] . It may be more feasible for the institutions to implement BYOD (Bring Your Own Device).

M-learning has become an interested topic of research in the countries with the existence of highly developed wireless network such as 3G [35] [36] and with the fast development of mobile technologies. As I stated earlier that M-learning is an extension of e-learning, therefore m-learning must benefits from the work done on e-learning to avoid wasting time and inventing things from scratch [33]. Moreover, interoperability issues between the two platform of e-learning and m-learning must be considered.

*A. Benefits of M-learning*

M-learning provides many benefits to the learning process and to the learner. Interaction: students are able to interact to each other and with their instructors easily [10]. Portability: mobile devices in general are portable and lighter than books. Students use PDAs to take their notes either by recording their voice, handwriting or by typing despite of their location [10]. Collaboration: regardless of their location, students are able to collaborate and work together on homework or even team projects [10]. Attractive: New age group are more attractive and engaged to the new technology represented by mobile devices such as PDAs, phones and game devices [10]. Raise motivation: learners via mobile devices are more engaged and interested to use and learn from it. "Bridging of the digital divide": The figures of mobile devices users are more than any other large systems due to their reasonable prices [10]. Instant learning: Raise their use and learning knowledge at any time. Assist disability: learning offer to all type of users including disables [10]. Many organizations prefer to equip the classroom with mobile devices other than desktop as it easier to do [37]. It may be more feasible for the institutions to implement BYOD (Bring Your Own Device).

Other advantages of M-Learning are learning unwittingly, life-long learning, learning adjusted to the learner time, place and situation [38]. Also, it eliminates communication barriers between learners and their instructors. Work effectively with student-centered learning. It support distance Learning (DL) and personalized learning.

*B. Limitation of M-learning*

Like any other type of learning and devices used, there are some limitations of M-learning Devices such as limited screens size of mobile phone and PDAs [10] [12]. Thus, it cause stress to the learner's eyes after long time of using the device, and the size allow only recap of information to be shown. Also, short battery life demands frequent charge [10][12] ,Inadequate memory capacity and slow to input text[12]. Moreover, limited wireless bandwidth is declined with bigger number of users. Also, it will not be easy to print except when it is connect to the network. As the technology change rapidly, devices might be inappropriate to provide and receive the service. Lack of cross-platform makes it hard to create contents for all types of devices [10]. Moving graphic is still hard to be used there even with the availability of 3G and 4G [38]. Higher level of destruction: a research conducted on two groups of students studying the same learning contents, the first using mobile devices while the second using desktop computers. Surprisingly, the latter group score higher grades than the first group as the first get distracted with doing other tasks in their mobile devices [39].

### III. CONTENT MANAGEMENT

The content management system (CMS) allows people to manage content from a central location. But in this paper we are not discussing about it. The main idea of this paper is to find how the content should be in the e-learning and m-learning tools. How it is affecting the success of E and M-learners? The Learning Management Systems (LMS) usually concentrate more on the organization of its content so that learners can access it easily. This is more related to the user friendliness of the E and M learning tools [40]. The LMS system will also try to learn the behavior of the learner and help to access the desirable content fast. Hence LMS is always better than CMS. In one of the research the researchers pointed out that if the content of the e-learning is not easily available to the learners then the e-learning project itself fails [41]. Hence for the successful e-learning project we have to include the learners also as a stakeholder and make sure that the e-learning system is more user-friendly [42]. In the recent days new LMS has more advanced options like open, social, personal, flexible, learning analytics and auto adjusting capability for mobile devices. This will benefit learners who are using e-learning as well as m-learning tools [45]. These kinds of LMS are called Integrated Learning Systems (ILS).

*A. E-learning content*

A common opinion about the e-learning among the management people is that e-learning designing cost is high as well as it is more complicated than the traditional face to face teaching and teaching material preparation. But once the e-learning content is designed and published online or in Intranet then the delivery cost is much cheaper than the conservative teaching and learning. Similarly the reusability of the materials is more only in the e-learning environment. The instructor-led and facilitated e-learning is becoming more famous now a days in the form of MOOC's (Massive Open Online Courses). This has all the advantages of traditional teaching and learning as well as we can take the advantage of e-learning system [46]. The synchronized e-learning system demands more time management from the students and this will groom the students with all the non-technical skills learned in the university system. The advantage of the e-learning is that all e-learning devices are having wide display compared to m-learning. Hence during the e-learning project either using CMS or using LMS we can keep all the related files in a folder with clear and obvious name. It will help the people to locate the content what they are looking for. The figure 1 shows how the content can be categorized and kept in the e-learning site [42].

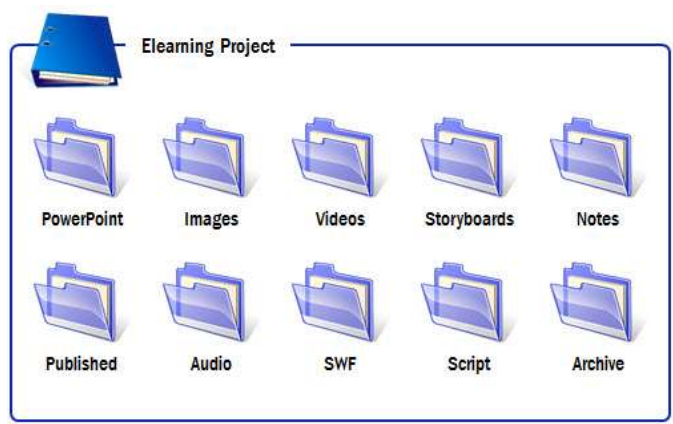

Fig. 1. Sample folder system in an e-learning tool

The e-learning content quality is very important. But in this paper we have assumed that the content is prepared by the subject expert. Being the technology expert we can put the content in the e-learning tool. But the very important thing to be noted is whether the end users are happy in the way the content is kept. Is it user friendly? Hence, after the content is designed by the subject expert as per the learning outcome and in proper order, it is the responsibility of the technical expert to keep the content in an easily accessible way to the end user. In e-learning environment the content can have more multimedia stuff and the story board creations are more easy and impressive for the end users. The options of moving between the menus will be easier in the e-learning system. Hence having more folders and organizing the content hierarchically will not spoil the user-friendliness of the e-learning system. E-learning content will be attractive even if we have only text and PowerPoint presentation kind of materials. The variety of contents can be kept in the e-learning system like storytelling, scenario based approach, toolkit approach, and demonstration-practice method. Hence e-learning design will not have much complication in having variety of materials, approaches in designing the content, organizing in many levels and still having user friendliness in it.

*B. M-learning content*

E-learning was dominating the ICT in education field for more than a decade. Since many institutes launched the e-learning system, it is not uncommon that they forget about further updations and maintenance. Even though it is having tracking facility about the students' performance, due to few drawbacks it was not revolutionized the ICT field in education. In recent days with the facilities what we have in the smartphones, popularized the m-learning concept. M-learning helps people to learn anywhere anytime as well as access the learning system just in time (JIT) [47]. But in mobile devices bookmark facility is not very effective. The great challenge in the m-learning is the content management. The content access by the user in JIT, content sequencing and distribution are not easy like e- learning. These challenges need to be resolved to make the m-learning as a silver bullet in ICT for education. The content accessing restriction on a subset of content is yet another great challenge in M-learning. Browsing from one topic to another topic in JIT is not yet made available for the learners in m-learning. The content type is also playing major role in the success of m-learning. If it is multimedia type, then it will be easier for the user to use it for learning. At the same time, the amount of memory and charge required to play that content is really taxing the learners. The internet data usage while using m-learning is also a big challenge. In a survey conducted in College of Applied Sciences Salalah, students prefer to use their mobile for m-learning when they are in some free Wi-Fi zone. Because the data package is costly and m-learning consumes lot of data even by using for a very short while. M-learning changing it faces by moving away from depending on the content management like a website than designing as apps. The great advantage of m-learning is that it supports BYOD (Bring Your Own Device) for learning. Always m-learning content should be designed for the device which has more limitations so that it will be working great with most of the average mobile devices. It is worthwhile to limit the content in the m-learning by giving only what the learner needs [48]. The content in m-learning cannot be like e-learning content, it should very short which can be read by the learner in 2 min maximum. If it has more content then it will be clumsier and make the end user tired by reading the content. In the recent MOOC's course designers have only video lectures with very less theory content. Also they give facility to download and read it at offline. Except the over use of memory, people prefer this method for their m-learning [49]. The learners are more comfortable with their mobile devices because they can do all in one device compare to the e-learning devices. Because of this all-in-one nature researchers are coining a term called handheld learning instead of m-learning [50]. The present LMS for m-learning using SCORM (shareable content object reference model) to share the learning objects among different LMSs.

### IV. COMPARISON OF E AND M LEARNING CONTENTS

The types of the content which should be in the e- learning and m-learning systems are different. The learners of the e-learning system may have more advantages in reading the text files, running a presentation and accessing e-books. But in the m-learning these contents may not be convenient for the learners. Hence multimedia contents are preferred by the m-learners [43]. On-demand LMS have the facility of accessing only desired content by the learners. This will be a great advantage for both E and M-learners. Moreover M-learners may have more advantage due to the limited storage, battery and internet access in their devices [44]. The new LMSs are having automatic adopting capability for both e-learning and m-learning [45]. So, when we use such systems then the content management system will be common for both the tools. Hence the design of the content management should not be biased for any one of the system. The content design will be little more complicated in the design stage. But, after launching the content online, managing the content will be easier. E-learning system can have heavy multimedia and graphical contents. If the internet connectivity is proper then accessing such content will be stress-free and convenient for the end users. It can be added that the e-learning system will be easier for the end users to move between the sections. Still m-learning system has the advantages over the e-learning system in accessing the content anywhere anytime and just in time. The apps design in m-learning which can work both dynamically and statically will become the future of new learning system using mobile devices. On-demand content options in apps make the apps more personalized than the e-learning system. If we make the same content design for e and m learning it will make both the systems to fail. Recent LMS systems and open source LMS are matured enough to provide a good e-learning system which meet the need of the institutions, teachers and learners. In recent days LMS companies are concentrating on their system solutions for mobile learning also. This has to be grown further to become a standard matured system for m-learning. Similarly the apps in mobile make a new trend in m-learning than using mobiles to read the e-learning contents. Based on our research outcome, we recommend few tips for the m-learning content designers:

- M-learning content should be accessed JIT, so that navigation will be easy.
- Buttons and links should be bigger than e-learning content
- M-learning content should be read in 2-4 minutes time (have one screen to deliver one specific topic)
- The data size of the content should not be large, give only the content, other added formatting will consume

huge memory and discourage people to use it due to data usage. Also it gets downloaded fast. (Bite-sized information)

V. CONCLUSION

The term e-learning is ruling the educational field for more than past three decades. Even after the transformation of this technology to m-learning, which is here for almost a decade, the need and demand for this technology is still green. These two technologies are equally used by the educational institutions and sometimes students can access the same content as per their wish and convenience. There are many advantages and disadvantages in both the systems. Most of the technical limitations are fixed by the growth in the technology. The success of these technologies now lies with the content management. In this paper we have observed that using the same content will not encourage the learners to use both the technologies simultaneously. This turned to be a reason for failure of the e or m learning projects. The content of the e-learning should be different from the content of m-learning. This paper also suggests the best way to have m-learning contents for the efficient use by the end users. E-learning technology became almost matured enough to give the learning material in an effective and efficient way. M-learning content delivery is becoming better due to apps design but still it has to go a long way to become a silver bullet for the learning solutions of the end users.